\documentclass[11pt]{article}
\usepackage{amsmath}
\usepackage{amssymb}
\usepackage{graphicx}
\usepackage[mathscr]{eucal}
\usepackage[usenames]{color}
\usepackage[latin5]{inputenc}
\usepackage{url}
\usepackage{authblk}

\numberwithin{equation}{section}

\newcommand{\be}{\begin{equation}}
\newcommand{\ee}{\end{equation}}

\begin{document}
\clearpage
\thispagestyle{empty}

\title{On the spectrum of the Schr\"odinger Hamiltonian of the one-dimensional conic oscillator perturbed by a point interaction}

\author[1,2,3]{S. Fassari\thanks{sifassari@gmail.com}}
\author[1]{M. Gadella\thanks{manuelgadella1@gmail.com}}
\author[4]{M.L. Glasser\thanks{laryg@tds.net}}
\author[1]{L.M. Nieto\thanks{luismiguel.nieto.calzada@uva.es}}
\affil[1]{Departamento de F\'{\i}sica Te\'{o}rica, At\'{o}mica y \'{O}ptica, and IMUVA, Universidad de Valladolid, 47011 Valladolid, Spain}
\affil[2]{CERFIM, PO Box 1132, Via F. Rusca 1, CH-6601 Locarno, Switzerland}
\affil[3]{Dipartimento di Fisica Nucleare, Subnucleare e delle Radiazioni, Universit\'a degli Studi Guglielmo Marconi, Via Plinio 44, I-00193 Rome, Italy}
\affil[4]{Department of Physics, Clarkson University, Potsdam, NY 13699, USA}

\date{\today}

\maketitle

\begin{abstract}
We decorate the one-dimensional conic oscillator $\frac{1}{2} \left[-\frac{d^{2} }{dx^{2} } + \left|x \right| \right]$ with a point impurity of either $\delta$-type, or local $\delta'$-type or even nonlocal $\delta'$-type. All the three cases are exactly solvable models,  which are explicitly solved and analysed, as a first step towards higher dimensional models of physical relevance.  We analyse the behaviour of the change in the energy levels when an interaction of the type $-\lambda\,\delta(x)$ or $-\lambda\,\delta(x-x_0)$ is switched on. In the first case, even energy levels (pertaining to antisymmetric bound states) remain invariant with $\lambda$ although odd energy levels (pertaining to symmetric bound states) decrease as $\lambda$ increases.  In the second, all energy levels decrease when the form factor $\lambda$ increases.   A similar study has been performed for the so called nonlocal $\delta'$ interaction, requiring a coupling constant renormalization, which implies the replacement of the form factor $\lambda$ by a renormalized form factor $\beta$. In terms of $\beta$, even energy levels are unchanged. However, we show the existence of level crossings: after a fixed value of $\beta$ the energy of each odd level, with the natural exception of the first one, becomes lower than the constant energy of the previous even level. Finally, we consider an interaction of the type $-a\delta(x)+b\delta'(x)$, and analyse in detail the discrete spectrum of the resulting self-adjoint Hamiltonian.

\end{abstract}



\section{Introduction}

One-dimensional point potentials in quantum mechanics \cite{AL,AK} and in quantum field theory (QFT) have received a lot of attention in the past two decades. First of all, they constitute a class of solvable or quasi-solvable potentials suitable to study basic quantum properties, stationary states, scattering, resonances, etc. In addition, they serve to model realistic physical situations with a number of practical applications. They are used to model several kinds of extra thin structures \cite{Z,Z1} or to point to defects in materials, so that effects like tunneling are easily studied. They are also used in the study of heterostructures, where they may appear in connection to an abrupt effective mass change \cite{GHNN}.

Given the increasing relevance of one-dimensional  quantum dots in the nanophysics literature \cite{HK,KV1,KV2}, point potentials can be used to model sharply peaked impurities inside the dot. In other physical contexts like scalar QFT on a line, point potentials serve to model impurities and provide external singular backgrounds where the bosons move \cite{MGM}. The spectra of Hamiltonians with $\delta$ and $\delta'$ point interactions provide one-particle states in scalar (1 + 1)-dimensional QFT systems \cite{AGM,AM,MKB}. In particular, configurations of two pure delta potentials added to the free Schr{\"o}dinger Hamiltonian have been used to describe scalar field fluctuations on external backgrounds \cite{GM}, as the corresponding scattering waves.  Delta point interactions allow implementation of some boundary conditions compatible with a scalar QFT defined on an interval \cite{BM}. The delta interaction is often multiplied by a real number $a$. In particular, this  coupling of $a$ to the $\delta$ potential mathematically describes the plasma frequencies in Barton's hydrodynamical model \cite{BAR}, characterizing the electro-magnetic properties of infinitely thin conducting  plates. On the other hand, the physical meaning of the $b$ coupling to the $\delta'$-interaction in the case of an interaction term of the form $b\delta'$ (with the prime denoting the derivative of the delta in the sense of distributions), appearing in the context of Casimir physics, has only recently been interpreted in \cite{BOR}: it describes the response of the orthogonal polarizability of a mono-atomically thin plate to the electromagnetic field.

Point interactions of the form $\delta$ or $\delta'$ or a linear combination thereof may be used as a perturbation of a free kinetic term like the Schr\"odinger  or the Salpeter free Hamiltonian \cite{AFR,EGU,WIE} or in combination with another potential. Examples of this potential are the harmonic oscillator \cite{GGN,FF,AFR1}, the constant electric field \cite{GGN}, the infinite square well \cite{GGN1,MA}, the conical oscillator \cite{CHINOS} that will be investigated in this article, or even the semi-oscillator which has been used as a simple example of potential showing resonance phenomena \cite{PIOTR,MAV}.  

There are physical reasons to consider the one-dimensional harmonic oscillator perturbed by point potentials, particularly in the theory of Bose-Einstein condensates \cite{UN,SH,GDB}. It has been very widely studied in the mathematical physics literature  \cite{FF,FI,FI1,PO,LUT,DEM,DEM1,COR,FIL} including the three-dimensional case \cite{FI2,BRU,AlbeverioFa16,AlbeverioFaRi16}.

In general, we have considered two methods to define point potentials. The former relies on the theory of self-adjoint extensions of symmetric operators with equal deficiency indices \cite{AL,AK,RS,KU}. Self-adjoint extensions of one-dimensional symmetric differential operators of first and second order have been extensively investigated. These extensions depend on one and four independent real parameters, respectively.  An example of a first order differential operator is given by the free Hamiltonian in the Dirac equation \cite{BRAS}, while that of a first order pseudodifferential operator by the free Hamiltonian in the Salpeter equation \cite{EGU,AK1}. Typically, the one-dimensional second order differential operator of interest in Physics is the free Schr\"odinger operator  $-1/2\,d^2/dx^2$ (with the choice $\hbar=m=1$), which has a four-parameter family of self-adjoint extensions \cite{KU}. These include point interactions like the Dirac delta $\delta$, its derivative $\delta'$ and others. Attempts to provide a physical meaning for all these extensions were made in \cite{KP}.  It is also important to remark that these extensions are defined by matching conditions at one or more points, namely the support of the point interactions, which define the domains of the self-adjoint extensions. As is well known, the support can even be a countable union of isolated points \cite{KU}, which makes the rigorous definition of Kronig-Penney type Hamiltonians possible \cite{LAR}.

The second method is based on the construction of the operator resolvent by either using the Krein formula \cite{AK} or by a calculation of the integral kernel, i.e. the  Green function, of the extension \cite{AFR,GGN,AFR1,GGN1,FI,FI1}. This is the method we are most frequently using in the present article, as will be explained later. 

In the case of the self-adjoint extensions of the Schr\"odinger operator, the most studied perturbations are those including $\delta$ and/or $\delta'$. Whilst there is no ambiguity as to the meaning of the $\delta$ interaction, there have been different interpretations of the meaning of the $\delta'$ interaction in the literature on the subject (see \cite{Lan} for a thorough critical review of the various intepretations found in the relevant literature). In this paper, we are going to compare the results, especially from the spectral point of view, obtained with two different choices of $\delta'$. Apart from the references given earlier on $\delta-\delta'$ interactions, we would like to include here some additional ones, not pretending to be exhaustive \cite{SEB,FAB,Z2,TN,ADK,FT,HC,FPP,Z3,GH,GNN,JUAN}. These references give us an idea of the interest on the subject.

The main objective of this paper is the study of the energy levels of the one-dimensional Hamiltonian inside which the standard harmonic confinement is replaced by the conic one, precisely
\begin{equation}\label{1}
H_0:=\frac12\left[ -\frac{d^2}{dx^2}+\,|x|\right],
\end{equation}
and decorated with a point potential of either the $\delta$ or $\delta'$ type. The interest of this model is similar to the harmonic oscillator decorated with the same kind of point potentials. In fact, the bound state spectra of this type of confining potentials has been a concern in quark confinement, see \cite{LEP}. Some particular aspects of the conic oscillator with point potentials have been considered elsewhere \cite{CHINOS,GN}.

 The point of departure is the resolvent of (\ref{1}) which can be easily written as an expansion in terms of its eigenfunctions and eigenvalues involving the Airy function, its derivative and their zeroes, as a consequence of the findings of \cite{LM}. Then, we decorate (\ref{1}) with a Dirac delta interaction.  First of all, we wish to anticipate at this stage that, differently from the notation conventionally adopted in the literature, the coupling constant will always be preceded by the minus sign throughout this note. The main reason for our choice is to show the manifestation of the Zeldovich  effect, also known as level rearrangement, for all the various models considered \cite{FF,AlbeverioFa16,Combescure,Farrell}. There is no difficulty in obtaining both the resolvent and the corresponding Green function of the resulting self-adjoint Hamiltonian. Bound states are obtained as solutions of a transcendental equation involving the Airy function and its derivative.  Next, we deal with the inverse problem: being given two energy values, and assuming the potential to be $\frac{1}{2}|x|-\lambda \delta(x-x_0)$, find the precise values of $\lambda$ and $x_0$. With such a generality, the problem has not a unique solution, differently from the result of \cite{FI1}, where only a local analysis is carried out for the lowest eigenvalues.

Next, we decorate (\ref{1}) with a $\delta'$ perturbation. There are two non-equivalent possibilities of defining this interaction and both correspond to two different matching conditions for the wave functions at the origin. These matching conditions determine different domains for $H_0$ and, therefore, different self-adjoint extensions of $H_0$, as defined on a restricted dense domain. 

The former, that we call the $|\delta'\rangle\langle \delta'|$ interaction, has been considered by several authors \cite{AL,AFR1,SEB} and is determined by the matching conditions: if  $\psi(x)$ is the wave function, its derivative $\psi'(x)$ is continuous at the origin but $\psi(x)$ itself satisfies the condition $\psi(0+)-\psi(0-)=-\beta\psi'(0)$, where $\beta$ is a fixed real number, i.e., the coefficient of the $\delta'$-interaction.  By using the standard matrix notation, the above condition can be written as:
\begin{equation}\label{2}
\left( \begin{array}{c}  \psi(0+) \\ [1ex]\psi'(0+)    \end{array}\right) =\left(\begin{array}{cc} 1 & -\beta \\ [1ex] 0 & 1  \end{array}\right)  \left( \begin{array}{c}  \psi(0-) \\ [1ex]\psi'(0-)    \end{array}\right)\,.
\end{equation}

  In this case, a naive calculation of the Green function shows divergencies, so that a renormalization is necessary. We perform first an ultraviolet energy cut-off and then the ensuing renormalization of the coupling constant. Renormalization procedures are necessary for point potentials not only in higher dimensions \cite{AL,AK}, but also in one-dimensional problems when the kinetic energy operator is proportional to the magnitude of the momentum and not to its square \cite{AFR,EGU,WIE}.  Again, eigenvalues are obtained through a transcendental equation  involving the Airy function and its derivative. The results are far more interesting than for the $\delta$ interaction: odd energy levels (pertaining to symmetric bound states, differently from the usual designation adopted for the harmonic oscillator) are not affected by the interaction. On the other hand, even energy levels (pertaining to antisymmetric bound states) are strongly affected. For negative values of the coefficient $\beta$, their energies are kept higher than the energy of the next lower odd level. However, after $\beta=0$ the energy decreases sharply and for a certain value of $\beta=\beta_0$, which is the same in all cases,  $\beta_0=1.37172$, the energy of each even energy level coincides with the energy of the next lower odd level. For values $\beta>\beta_0$, the energy of the even level decreases further, so that we are in the presence of a {\it quantum  phase transition}.

A second $\delta'$ interaction, which is compatible with the $\delta$ interaction, so that we may compose interactions of the form $-a\delta(x)+b\delta'(x)$ \cite{GNN,JUAN,GOL}, comes form the consideration of the following matching conditions for the wave function at the origin:
\begin{equation}\label{3}
\left( \begin{array}{c}  \psi(0+) \\ [1ex]\psi'(0+)    \end{array}\right) =\left(\begin{array}{cc} \displaystyle\frac{1+b}{1-b} & 0 \\[2ex] \displaystyle\frac{2a}{1-b^2} & \displaystyle\frac{1-b}{1+b}  \end{array}\right)  \left( \begin{array}{c}  \psi(0-) \\ [1ex]\psi'(0-)    \end{array}\right)\,.
\end{equation}
The problem is explicitly solved and our findings show that, if the potential is purely of $\delta'$ type ($a=0$), there are no new eigenvalues. On the other hand, if the potential is purely of $\delta$ type ($b=0$), we get the same results of Section~3.

\section{Some mathematical preliminaries}

The strictly positive operator
\begin{equation}\label{(1.1)}
H_{0}= \frac{1}{2} \left[-\frac{d^{2} }{dx^{2} } + \left|x \right| \right]=\sum _{n=1}^{\infty }E _{n}\left|\psi _{n}\right\rangle\left\langle\psi _{n}\right|,
\end{equation}
acting on $L^{2} (-\infty,+\infty)$, whose eigenfunctions and eigenvalues have been thoroughly investigated in \cite{LM} (Theorem 3.5 and Corollary 3.6), is characterized by having a compact resolvent, whose integral kernel is given by:
\begin{eqnarray}
\label{(1.2)}
(H_{0}-E)^{-1}(x,y) &=&\sum _{n=1}^{\infty }\frac{\psi _{n}(x)\psi _{n}(y)}{E _{n}-E}=-\frac{Ai(x_{>}-2E)Ai(-x_{<}-2E)}{Ai(-2E)Ai'(-2E)} \\ [1ex]
&=&
-\frac{Ai\left(\displaystyle\frac{x+y+|x-y|}{2}-2E\right)\, Ai\left(\displaystyle\frac{-x-y+|x-y|}{2}-2E\right)}{Ai(-2E)Ai'(-2E)},
\nonumber
\end{eqnarray}
for any  $E$ in the resolvent set $\rho(H_{0})$, taking advantage of (3.8) in \cite{GN}. Before moving forward, it is crucial to point out that in the case of the Hamiltonian $H_{0}$, differently from the standard notation used for the harmonic oscillator Hamiltonian, the eigenvalues and the eigenfunctions of the symmetric (resp. antisymmetric) bound states are labelled by an odd (even) index.

We remind the reader that a positive compact operator on a given Hilbert space is said to belong to the Schatten class of index $\gamma$ if the sequence of the $\gamma$-th powers of its eigenvalues is summable. A compact operator  belongs to the Schatten class of index $\gamma$ if and only if its modulus does. Hence, the following theorem can be stated.
\smallskip

\noindent
\textbf{Theorem 2.1} The resolvent of the operator $H_{0}$ is a compact operator on $L^{2} (-\infty,+\infty)$ belonging to the Schatten class of index $3/2+\epsilon$.
\smallskip

\noindent \textbf{Proof.} As a consequence of the first resolvent identity, it is sufficient to prove the statement only for any $E<0$. By taking account of the findings of \cite{LM}, implying that  $E_{2n}<E_{2n+1}$, we get for any $E<0$:
\begin{eqnarray}\nonumber
\sum _{n=1}^{\infty }\frac{1}{(E_{n}-E) ^{3/2+\epsilon}} &<& \sum _{n=1}^{\infty }\frac{1}{E _{2n-1} ^{3/2+\epsilon}}+\sum _{n=1}^{\infty }\frac{1}{E _{2n}^{3/2+\epsilon}}
 <\frac{1}{E _{1} ^{3/2+\epsilon}}+2\sum _{n=1}^{\infty }\frac{1}{E _{2n}^{3/2+\epsilon}} \\ [1ex]
 &<& \frac{1}{E_{1} ^{3/2+\epsilon}}+\frac{2^{11/2+3\epsilon}}{(3\pi) ^{1+\frac{2\epsilon}{3}}}\sum _{n=1}^{\infty }\frac{1}{(4n-1) ^{1+\frac{2\epsilon}{3}}}<\infty.
\end{eqnarray}

\noindent
\textbf{Remark 2.2} It might be worth pointing out at this stage that the resolvent of the Hamiltonian of the quantum harmonic oscillator is instead a compact operator belonging to the Schatten class of index $1+\epsilon$. 
\smallskip

Another result, playing a crucial role in the following, comes next.
\smallskip

\noindent
\textbf{Theorem 2.3} The function $(H_{0} -E) ^{-1/2}(\cdot ,x_{0} )$ is square integrable for any real $x_{0}$ and any $E$ in the resolvent set of  $H_{0}$. 
\smallskip

\noindent \textbf{Proof.} As a result of (2.2) and the smoothness of the Airy function and its derivative, we get:
\begin{eqnarray}\nonumber
\parallel(H_{0}-E) ^{-1/2}(\cdot,x_{0})\parallel^{2}_{2} =\sum _{n=1}^{\infty }\frac{\psi_{n}^{2} (x_{0})}{E_{n}-E}=(H_{0} -E) ^{-1}(x_{0},x_{0} )
\\ [1ex]
=-\frac{Ai(x_{0}-2E)Ai(-x_{0}-2E)}{Ai(-2E)Ai'(-2E)}<\infty,
\end{eqnarray}
which fully proves our claim.
\smallskip

\noindent
\textbf{Remark 2.4} It might be worth reminding the reader that the corresponding sequence ${\psi_{n}^{2}(x_{0})}/(n+1/2)$ for the Hamiltonian of the quantum harmonic oscillator behaves asymptotically like $n^{-3/2}$ (see, e.g., \cite{FF,AFR1,FI,FI1,Mit1,Mit2}).

\section{The spectrum of $H_{0}$  perturbed by an  attractive $\delta$-interaction}

By noticing that
\begin{eqnarray}\label{(2.1)}
(H_{0}-E)^{-1/2}\left|\delta(x-x_{0}) \right\rangle\left\langle \delta(x-x_{0})\right| (H_{0}-E)^{-1/2}  =(H_{0}-E)^{-1/2}\delta(x-x_{0})(H_{0}-E)^{-1/2} = \nonumber  \\  [1ex]
 =\left| (H_{0} -E) ^{-1/2} (\cdot ,x_{0} ) \right\rangle\left\langle (H_{0} -E)^{-1/2} (x_{0} ,\cdot )\right|,
\end{eqnarray}
the resolvent of 
\be\label{(2.2)}
H_{\lambda,x_{0}}=H_{0}-\lambda\delta(x-x_{0}),\quad \lambda >0
\ee
can be written as 
\begin{eqnarray}\label{(2.3)}
&& (H_{\lambda,x_{0}}-E)^{-1}= \\
&& \qquad =(H_{0}-E)^{-1/2} \left[1-\lambda\left| (H_{0} -E) ^{-1/2} (\cdot ,x_{0} ) \right\rangle\left\langle (H_{0} -E)^{-1/2} (x_{0} ,\cdot )\right|\right]^{-1}(H_{0}-E)^{-1/2} , \nonumber
\end{eqnarray}
for any $E<0$ such that
\be
\lambda\parallel(H_{0} -E) ^{-1/2}(\cdot ,x_{0} )\parallel^{2}_{2}=\lambda\sum _{n=1}^{\infty }\frac{\psi _{n}^{2} (x_{0})}{E_{n}-E}=\lambda(H_{0} -E) ^{-1}(x_{0},x_{0} )<1.
\ee
Furthermore, since
\begin{eqnarray}
&& \left[\left| (H_{0} -E) ^{-1/2} (\cdot ,x_{0} ) \right\rangle\left\langle (H_{0} -E)^{-1/2} (x_{0} ,\cdot )\right|\right]^{n} = \nonumber \\
&& \qquad  =\left[\parallel(H_{0} -E) ^{-1/2}(\cdot ,x_{0} )\parallel^{2}_{2}\right]^{n-1}\left| (H_{0} -E) ^{-1/2} (\cdot ,x_{0} ) \right\rangle\left\langle (H_{0} -E)^{-1/2} (x_{0} ,\cdot )\right| \nonumber \\
&& \qquad  =\left[(H_{0} -E) ^{-1}(x_{0},x_{0} )\right]^{n-1}\left| (H_{0} -E) ^{-1/2} (\cdot ,x_{0} ) \right\rangle\left\langle (H_{0} -E)^{-1/2} (x_{0} ,\cdot )\right|,
\end{eqnarray}
the operator in the middle of the right hand side of (\ref{(2.3)}) can be written as
\begin{eqnarray}
\nonumber
1+\frac{\lambda\left| (H_{0} -E) ^{-1/2} (\cdot ,x_{0} ) \right\rangle\left\langle (H_{0} -E)^{-1/2} (x_{0} ,\cdot )\right|}{1-\lambda(H_{0} -E) ^{-1}(x_{0},x_{0} ) } =
1+\frac{\left| (H_{0} -E) ^{-1/2} (\cdot ,x_{0} ) \right\rangle\left\langle (H_{0} -E)^{-1/2} (x_{0} ,\cdot )\right|}{\frac{1}{\lambda}-(H_{0} -E) ^{-1}(x_{0},x_{0} ) }.
\label{(2.5)}
\end{eqnarray}
Hence, we get the following result.
\smallskip

\noindent
\textbf{Theorem 3.1} The self-adjoint operator making rigorous mathematical sense of the heuristic expression (\ref{(2.2)}) is the one whose resolvent is given for any  $E$ in the resolvent set $\rho(H_{0})$  by:
\be\label{(2.6)}
(H_{\lambda,x_{0}}-E)^{-1}=(H_{0} -E) ^{-1}+\frac{\left| (H_{0} -E) ^{-1} (\cdot ,x_{0} ) \right\rangle\left\langle (H_{0} -E)^{-1} (x_{0} ,\cdot )\right|}{\frac{1}{\lambda}-(H_{0} -E) ^{-1}(x_{0},x_{0} )}.
\ee
Furthermore, $H_{\lambda,x_{0}}$ regarded as a function of $\lambda$ is an analytic family in the sense of Kato. 
\smallskip

Given that the eigenvalues of the operator $H_{\lambda,x_{0}}$ are exactly the poles of its resolvent, in order to achieve an accurate description of the spectrum (obviously being exclusively discrete) of the operator we need only look for the solutions of the equation:
\be\label{(2.7)}
\frac{1}{\lambda}-(H_{0} -E) ^{-1}(x_{0},x_{0} )=0.
\ee
By inserting the explicit expression for the resolvent of $H_{0}$ provided in \cite{GN}, (\ref{(2.7)}) can be recast as:
\be\label{(2.8)}
\lambda=-\frac{Ai(-2E)Ai'(-2E)}{Ai(x_{0}-2E)Ai(-x_{0}-2E)}.
\ee
\smallskip

\noindent
\textbf{Remark 3.2} A remarkable feature of the above bound state equation is that its simple expression in terms of Airy functions holds for any real $x_{0}$, differently from its analogue for the harmonic oscillator perturbed by a Dirac distribution that can be expressed as a ratio of Gamma functions only for $x_{0}=0$ (see, e.g., \cite{FF,AFR1,FI,FI1,PO}).

\begin{figure}[ht]
\centering
\includegraphics[width=8cm]{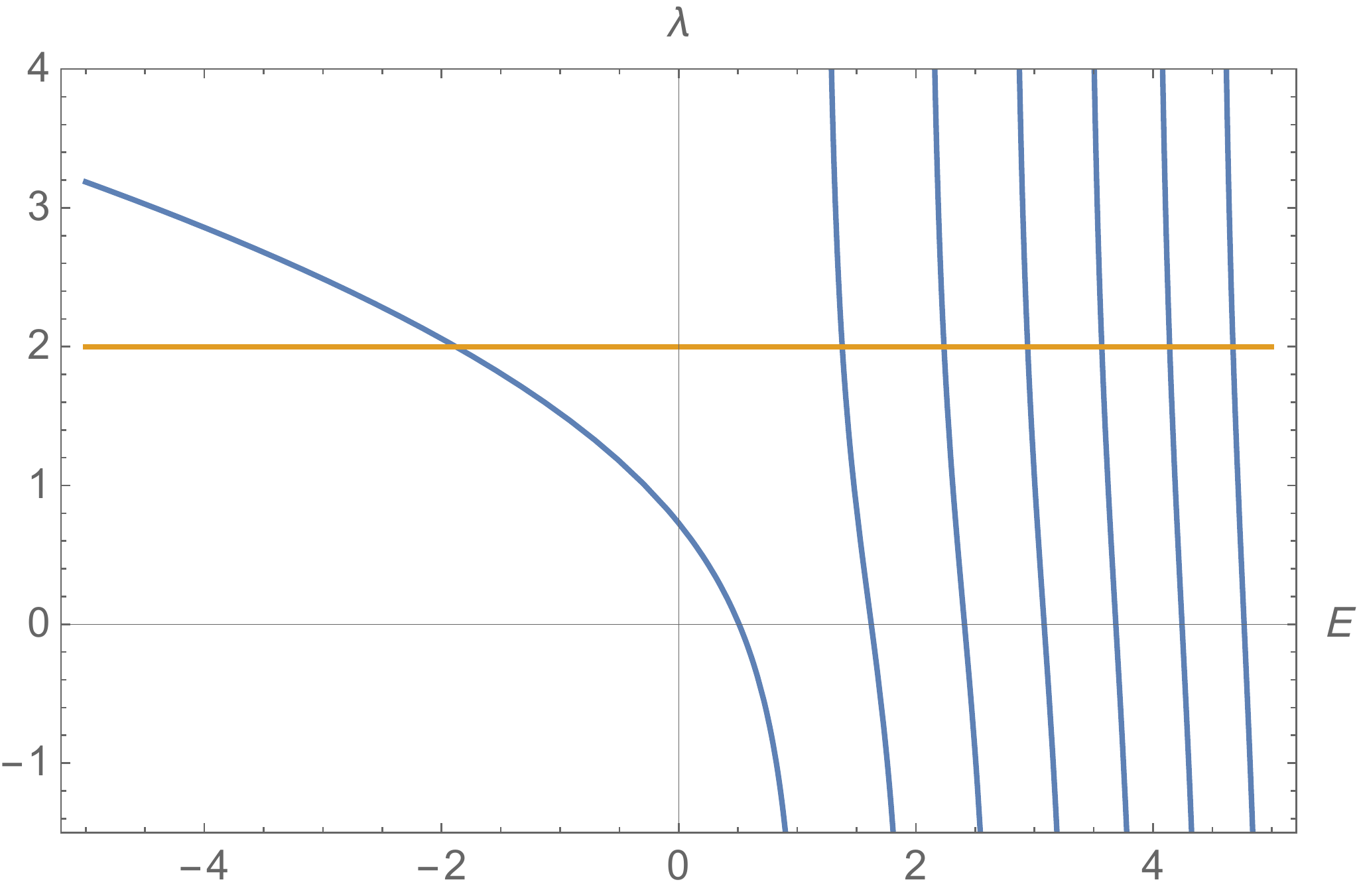}
\caption{\small  The  $n$-th eigenvalue of $H_{\lambda=2,x_{0}=0}$ is given by the intersection between the $n$-th branch of the right hand side of (\ref{(2.9)}) as a function of the energy and the horizontal line $\lambda=2$.
\label{fig1}}
\end{figure}

Let us start the analysis of the latter equation by considering the special case $x_{0}=0$. The above equation becomes:
\be\label{(2.9)}
\lambda=-\frac{Ai'(-2E)}{Ai(-2E)}.
\ee
By plotting the right hand side of (\ref{(2.9)}) as a function of the energy, the $n$-th eigenvalue is determined by the intersection between the $n$-th branch of the function and the horizontal line pertaining to the given value of the coupling constant, as shown in Figure~\ref{fig1} for $\lambda=2$.

Since $\lambda(E,x_{0}=0)$ consists of infinitely many branches, each of which is a strictly decreasing function of the energy parameter, the inverse function of the $n$-th branch is nothing else but the $(2n-1)$-th eigenenergy $E_{2n-1}(\lambda,x_{0}=0)$. The eigenvalues pertaining to the antisymmetric bound states are not modified at all by the perturbation so that, for any real $\lambda$,  $E_{2n}(\lambda,x_{0}=0)=E_{2n}(0,x_{0}=0)=E_{2n}$ obtained from (\ref{(1.2)}) as the solutions of $Ai(-2E)=0$.

\begin{figure}[ht]
\centering
\includegraphics[width=7cm]{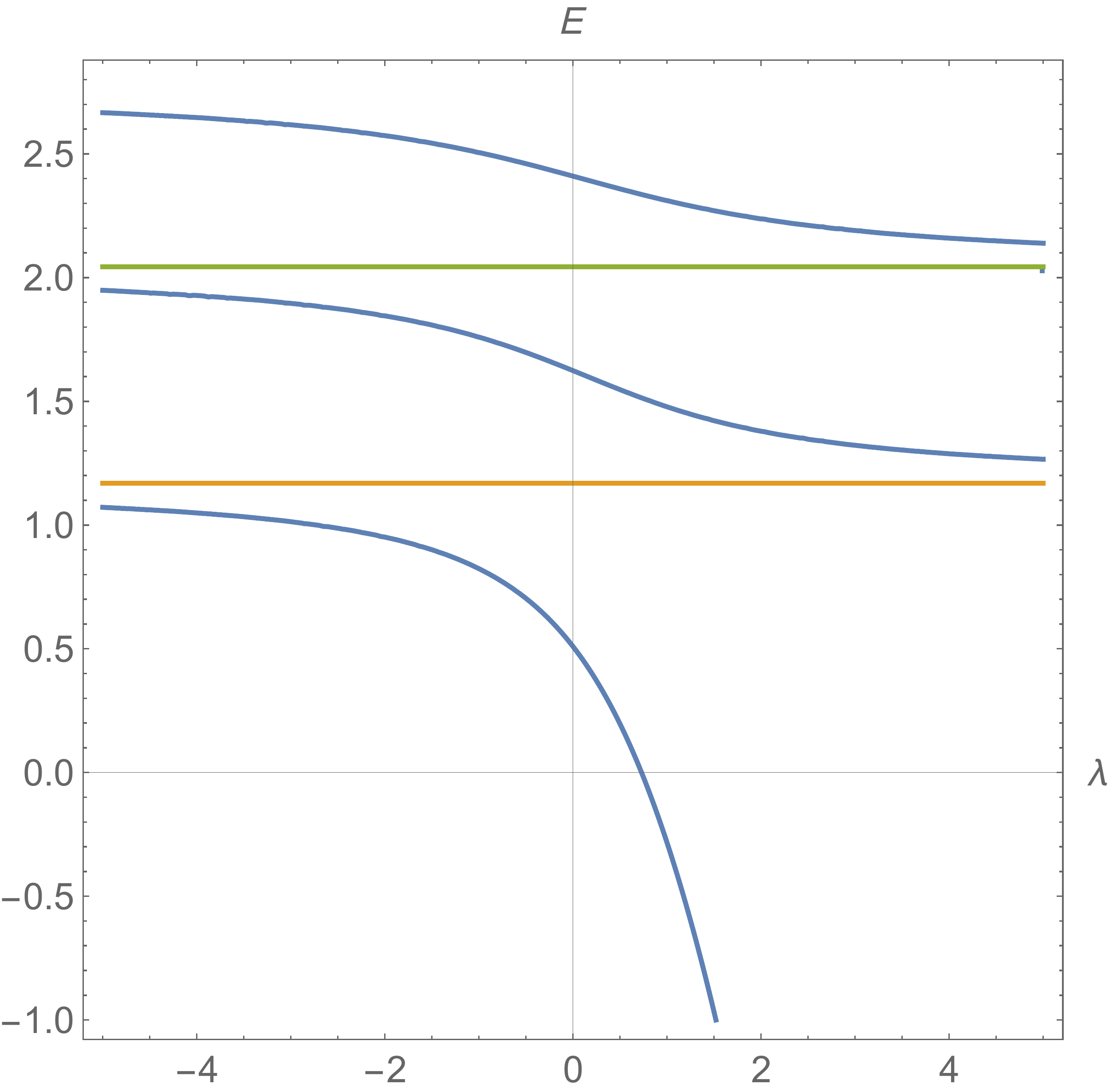}
\caption{\small Plot of the five lowest eigenenergies $E_{n}(\lambda,x_{0}=0)$. The even eigenvalues do not depend on $\lambda$: $E_2(\lambda,0)= E_2(0,0)$ in yellow and $E_4(\lambda,0)= E_4(0,0)$ in green.
\label{fig2}}
\end{figure}

The five lowest eigenvalues as functions of the coupling constant are plotted in Figure~\ref{fig2}. The eigenvalues of the two lowest antisymmetric bound states, namely $E_{2}(0,x_{0}=0),E_{4}(0,x_{0}=0)$, are the horizontal asymptotes of those of the symmetric bound states.

With reference to Figure 1 in \cite{CHINOS}, two important differences are to be pointed out. First of all, the unperturbed operator in \cite{CHINOS} is twice our free Hamiltonian. Moreover, our Hamiltonian $H_{\lambda,x_{0}}$ is characterized by the presence of the minus sign in front of the coupling constant, thus making the interaction more attractive moving from the left to the right along the horizontal axis. We have chosen to do so since we wish to show the manifestation of the Zeldovich effect, also known as level rearrangement, for the model being analysed here, given that the spectrum of the harmonic oscillator perturbed by a Dirac distribution (in both one and three dimensions) exhibits the level rearrangement pattern, as shown in \cite{FF,AlbeverioFa16} (see also \cite{Combescure,Farrell} for additional information on the Zeldovich effect for Hamiltonians with rapidly decaying potentials and the definition of the Zeldovich spiral).

\begin{figure}[ht]
\centering
\includegraphics[width=8cm]{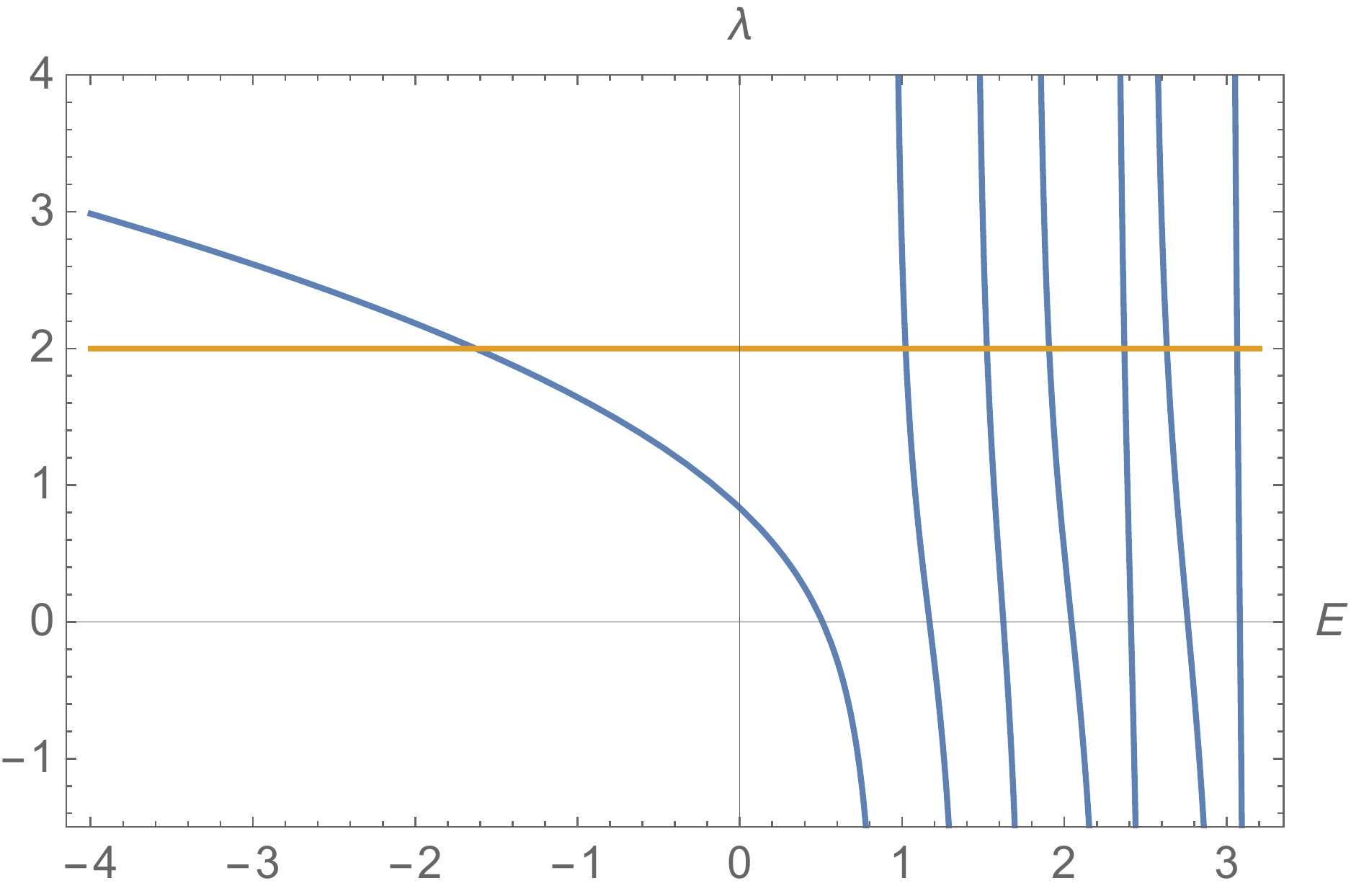}
\caption{\small The $n$-th eigenvalue of $H_{\lambda=2,x_{0}=1/2}$ is given by the intersection between the $n$-th branch of the right hand side of (\ref{(2.8)})  as a function of the energy and the horizontal line $\lambda=2$.
\label{fig3}}
\end{figure}

\begin{figure}[ht]
\centering
\includegraphics[width=7cm]{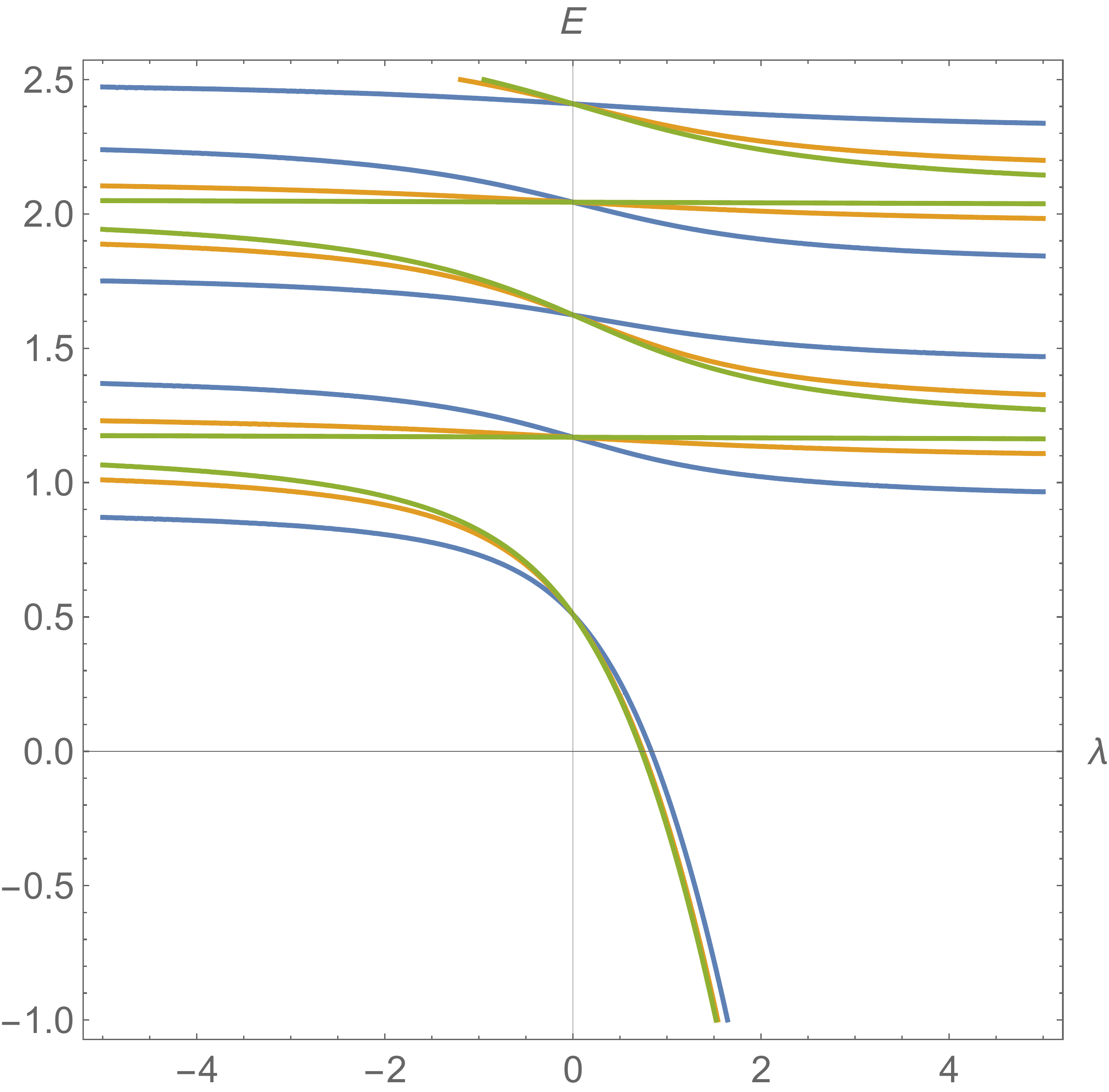}
\caption{\small Plot of the five lowest eigenenergies $E_{n}(\lambda,x_{0}),n=1,...,5$, for $x_0=0.5$ (blue), $0.2$ (yellow) and $0.05$ (green).
\label{fig4}}
\end{figure}

As can be expected on the basis of the smooth dependence of the right hand side of (\ref{(2.8)}) on $x_{0}$, the spectral stucture of the odd eigenvalues (pertaining to the symmetric bound states) does not undergo any major qualitative change. However, the antisymmetric bound states are also affected by the perturbation, so that the even eigenvalues are no longer given by those of $H_{0}$ but by functions of the coupling $\lambda$. Whilst the analogue of Figure~\ref{fig1} is plotted in Figure~\ref{fig3}, the plot of the five lowest eigenenergies as functions of $\lambda$ for $x_{0}=0.05,0.2,0.5$ is shown in Figure~\ref{fig4}.

\begin{figure}[ht]
\centering
\includegraphics[width=8cm, height=6cm, keepaspectratio]{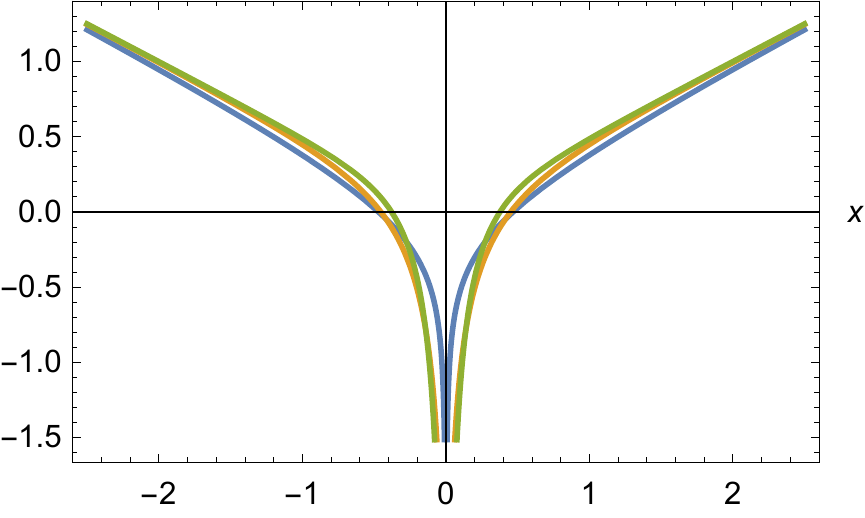}
\caption{\small Plot of the potentials $\frac{1}{2}|x|-V_{n}(x)$, for $n=1$ (blue), $5$ (yellow) and $10$ (green).
\label{fig5}}
\end{figure}

\noindent
\textbf{Remark 3.3} Although we are not going to formally state and prove the analogues of Theorem 2.2 in \cite{FI} and Theorem 2.1 in \cite{ FI1}, it is almost evident that $H_{\lambda,x_{0}}$ is the norm resolvent limit of any Hamiltonian with a potential given by the sum of $\frac{1}{2}|x|$ and a sequence of sharply peaked attractive potentials $V_{n}(x)$ approximating the Dirac distribution. In Figure~\ref{fig5} we show an example of such a combination for $\lambda=1,x_{0}=0$ and $V_{n}(x)=\frac{n}{3}\frac{e^{-|nx|^\frac{2}{3}}}{|nx|^\frac{1}{3}}$, an example of a funnel shaped potential.

\subsection{The inverse problem: identifying the potential from measured values of the energy}

If one is given two energy levels $E_1$, and $E_2$, the system can be ``designed", that is an impurity location and strength can be specified, which will reproduce these  two energy values. From (\ref{(2.8)}), one has only to solve the equations
\begin{eqnarray}
\label{larry1}
\frac{Ai(-2E_1)Ai'(-2E_1)}{Ai(-2E_2)Ai'(-2E_2)}=\frac{Ai(x_0-2E_1)Ai(-x_0-2E_1)}{Ai(x_0-2E_2)Ai(-x_0-2E_2)}.
\\
\label{larry2}
\lambda=-\frac{Ai(-2E_1)Ai'(-2E_1)}{Ai(x_0-2E_1)Ai(-x_0-2E_1)}.
\end{eqnarray}

We have been motivated to investigate this issue since a similar problem for the case of a delta impurity in a harmonic oscillator well was dealt with by Fassari and Inglese \cite{FI1} who, however, had restricted their analysis to the case when $E_1$ and $E_2$ are the two lowest eigenvalues, due to their use of the specific integral expressions derived from the so-called Mehler kernel and not the global ones in terms of parabolic cylindrical functions. They showed that there is a unique solution  and they examined its stability with respect to small variations in the energy values.

\begin{figure}[ht]
\centering
\includegraphics[width=8cm]{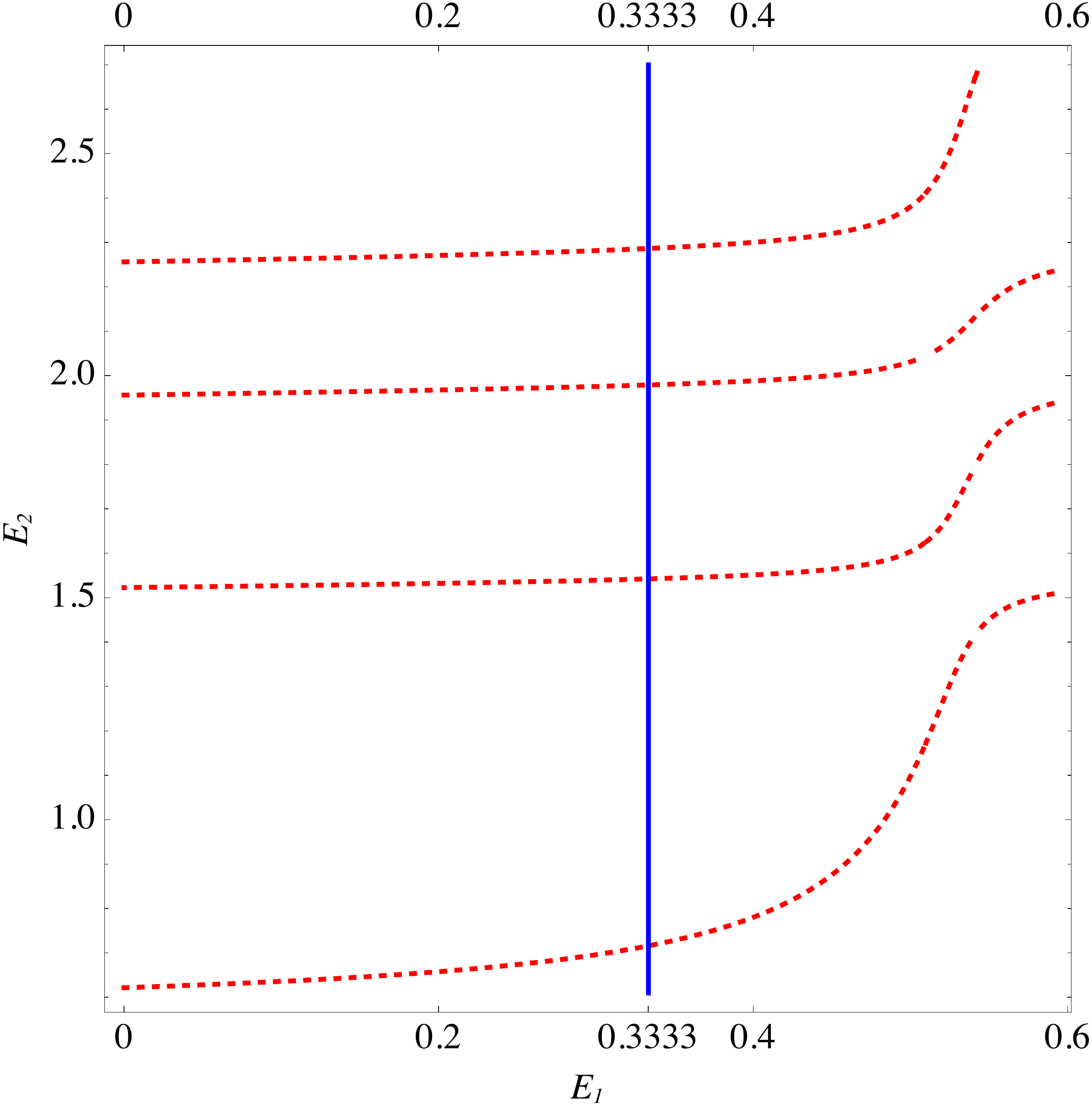}
\caption{\small Plot of the implicit equation (\ref{larry1}) for $x_0=1.2557$, which has an infinity of branches (dotted red lines). For a fixed value of $E_1$, let us say $E_1=0.3333$ (blue line), the infinite solutions of the inverse problem are obtained.
\label{figinverse}}
\end{figure}

In the present case equations (\ref{larry1})-(\ref{larry2}) are easily solved graphically and, since the Airy functions are oscillatory, will produce infinitely many potentials possessing the given levels. For example, let us  choose $x_0=1.2557$ and plot the pair of energies  $E_1,E_2$ which the delta potential at this point  will possess (each will correspond to a unique value of $\lambda$, of course.) We illustrate this in Figure~\ref{figinverse} The compatible energy pairs lie along the dotted lines, one being $(E_1,E_2)=(0.3333,0.7158)$ with $\lambda=1.3602$, but with other possibilities as $(E_1,E_2)=(0.3333,1.5423)$, $(E_1,E_2)=(0.3333,1.9791)$, etc.

\section{The spectrum of $H_{0}$  perturbed by an  attractive $\delta'$-interaction}

In this section we are going to analyse the results that two different interpretations of the $\delta'$-interaction give to the problem of dressing the conic oscillator with a point perturbation. We will first consider the approach based on \cite{AFR1} (see also \cite{AL,Albeverio16}) and then the one based instead on \cite{GGN,GGN1}.

\subsection{Nonlocal $\delta'$ interaction: operator resolvent approach}

First of all, in view of what will be developed throughout this section, it is of the utmost importance to realise that, for any Schwartz function $f$, we have 
$$
(f,\delta' f)\neq(f,\delta')(\delta', f)=|f'(0)|^{2}. 
$$
Hence, the counterpart of the first equality in (\ref{(2.1)}) will no longer hold in the case of the derivative of the Dirac distribution. As a consequence, we wish to stress at this stage that in the following we are going to deal with the singular perturbation $\left|\delta'(x) \right\rangle\left\langle \delta'(x)\right|$ and not with the $\delta'$-potential, in analogy with what was done in \cite{AFR1} for the model with the harmonic confinement. 

 Given that
\begin{eqnarray}
(H_{0}-E)^{-1/2}\left|\delta'(x) \right\rangle\left\langle \delta'(x)\right| (H_{0}-E)^{-1/2} \!\!&\!=\!&\!\! \left| \sum _{n=1}^{\infty }\frac{\psi' _{n}(0)\psi _{n}}{(E _{n}-E)^{1/2}} \right\rangle\left\langle  \sum _{n=1}^{\infty }\frac{\psi' _{n}(0)\psi _{n}}{(E _{n}-E)^{1/2}}\right| \nonumber \\
\!\!&\!=\!&\!\! \frac{1}{2}\left| \sum _{n=1}^{\infty }\frac{\psi _{2n}}{(E _{2n}-E)^{1/2}} \right\rangle\left\langle  \sum _{n=1}^{\infty }\frac{\psi _{2n}}{(E _{2n}-E)^{1/2}}\right|, 
\label{(5.1)}
\end{eqnarray}
it is almost immediate to realize that the sequence of the square of the coefficients behaves asymptotically like $n^{-2/3}$ (see again Theorem 3.5 and Corollary 3.6 in \cite{LM}) implying the divergence of its series. As was done in \cite{AFR1} in the presence of the harmonic confinement, a possible strategy leading to the rigorous definition of the self-adjoint Hamiltonian is the introduction of an energy cut-off and the ensuing renormalization of the coupling constant so that a suitable operator may be obtained in the norm resolvent limit as the cut-off gets removed.

The counterpart of the function defined in (2.3) in \cite{AFR1} will be given by:
\be\label{(4.2)}
\Psi (x;E):=\frac{1}{2^{1/2}} \sum _{n=1}^{\infty }\frac{\psi _{2n}(x)}{(E_{2n}-E)},
\ee
with
\be\label{(4.3)}
||\Psi (E)||^{2} _{2}=\frac{1}{2} \sum _{n=1}^{\infty }\frac{1}{(E _{2n}-E)^{2}}<\infty,
\ee
since the sequence with terms $(E _{2n}-E)^{-2}$ behaves asymptotically like $n^{-4/3}$.

\begin{figure}[ht]
\centering
\includegraphics[width=8cm]{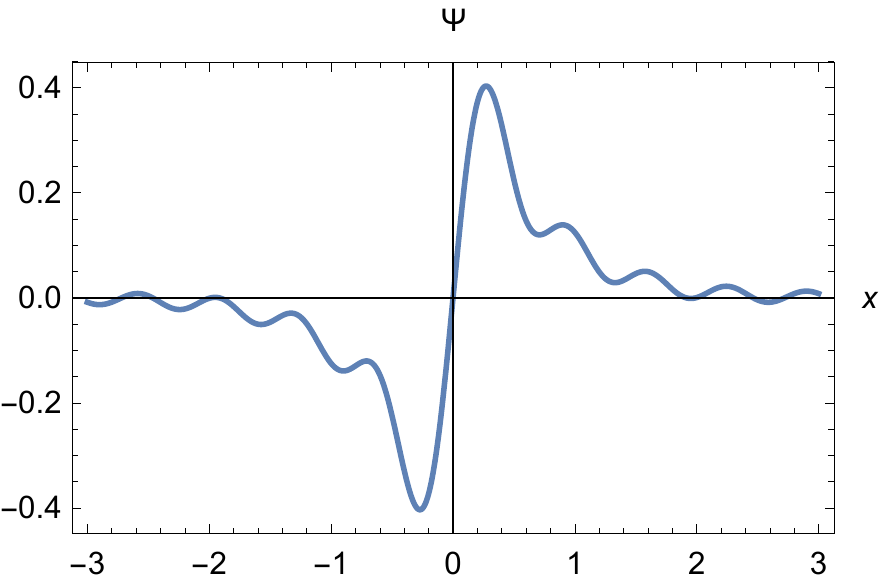}
\caption{\small The approximation of $\Psi(x;E)$ obtained by truncating its series at the 200th term.
\label{fig7}}
\end{figure}

The plot of the approximation of $\Psi (x;E)$ obtained by taking only the first 200 terms of its series is shown in Figure~\ref{fig7}. The graph clearly shows the action of the attractive $\delta'$-interaction: the continuity of the wave function at the origin is bound to be lost, given that  $\Psi (0_{+};E)-\Psi (0_{-};E)=-\beta\Psi' (0;E)$.

The counterpart of (2.4a) defining the coupling constant being renormalized will be given by:
\be\label{(4.4)}
\frac{1}{\mu _{\beta}(N)}= \frac{1}{\beta}+\frac{1}{2} \sum _{n=1}^{N }\frac{1}{E _{2n}},
\ee
where the integer $N$ defines the ultraviolet energy cut-off.

As the intermediate steps would be identical to those of \cite{AFR1}, the only difference being the meaning of the free Hamiltonian $H _{0}$, we omit them and write directly the limit as  $N\rightarrow \infty$  in the norm topology of the resolvents, that is to say the counterpart of (2.7) in \cite{AFR1}:
\begin{eqnarray}
R(\beta,E)&=&(H_{0} -E) ^{-1}+ \frac{\left|\Psi (E) \right\rangle\left\langle\Psi (E)\right|}{\beta^{-1 }+\frac{1}{2}\left[ \displaystyle\sum _{n=1}^{\infty }\frac{1}{E _{2n}}-\sum _{n=1}^{\infty }\frac{1}{(E_{2n}-E)}\right]} \nonumber \\
&=&(H_{0} -E) ^{-1}+ \frac{\left|\Psi (E) \right\rangle\left\langle\Psi (E)\right|}{\displaystyle \beta^{-1 }-\frac{E}{2} \sum _{n=1}^{\infty }\frac{1}{E _{2n}(E_{2n}-E)}}.
\label{(4.5)}
\end{eqnarray}
The series in the denominator of the second term on the right hand side of (\ref{(4.5)}) is convergent since the sequence 
${1}/(E_{2n}(E_{2n}-E))$ behaves asymptotically like $n^{-4/3}$.

Therefore, $R(\beta,E)$, being the sum of the free resolvent and an operator of rank one, is a perfectly defined bounded operator. The ensuing task would be to show that the latter operator is indeed the resolvent of a self-adjoint operator, which would require a rather lengthy and technically detailed proof. However, such a proof will be omitted here given that it would basically mimic the analogous one for the negative Laplacian perturbed by a Dirac distribution in three dimensions fully provided in \cite{AL} or the one for the one-dimensional Salpeter Hamiltonian perturbed by a Dirac distribution in \cite{AFR}. 
Hence, the following result can be stated.
\smallskip

\noindent
\textbf{Theorem 4.1} The self-adjoint operator $H_{\beta}$ making rigorous mathematical sense of the heuristic expression  $H_{0}-\mu(\beta)\left|\delta'(x) \right\rangle\left\langle \delta'(x)\right|,$ with $\mu(\beta)= {2\beta}/(2+\beta \sum_{n=1}^{\infty}E^{-1}_{2n})$, is the one whose resolvent is given for any  $E$ in the resolvent set $\rho(H_{0})$ by:
\be\label{(4.6)}
(H_{\beta} -E) ^{-1}=(H_{0} -E) ^{-1}+ \frac{\left|\Psi (E) \right\rangle\left\langle\Psi (E)\right|}{\displaystyle \beta^{-1 }-\frac{E}{2} \sum _{n=1}^{\infty }\frac{1}{E_{2n}(E_{2n}-E)}}.
\ee
Furthermore, $H_{\beta}$ regarded as a function of $\beta$ is an analytic family in the sense of Kato.
\smallskip

Given that the eigenvalues of the operator $H_{\beta}$ are exactly the poles of its resolvent, in order to achieve an accurate description of the spectrum (obviously being exclusively discrete) of the operator, we need only look for the solutions of the equation:
\be\label{(4.7)}
\beta^{-1 }=\frac{E}{2} \sum _{n=1}^{\infty }\frac{1}{E _{2n}(E_{2n}-E)}.
\ee

In analogy with what was done in \cite{AFR1}  for the harmonic oscillator, if one could recast the series on the right hand side of (\ref{(4.7)}) into an expression involving Airy functions and their derivatives, the detailed study of the above bound state equation would be more accessible. The goal can actually be achieved by taking account of (1.3) in \cite{AK}  (see also \cite{Grosse}), so that:
\begin{eqnarray}
\sum _{n=1}^{\infty }\frac{1/2}{(E_{2n}-E)}
\!\!&\!=\!&\!\!
(\delta',\sum _{n=1}^{\infty }\frac{\left|\psi_{n} \right\rangle\left\langle\psi_{n}\right|}{E _{n}-E}\delta') = 
\int _{-\infty }^{\infty }\int _{-\infty }^{\infty}\delta'(x)\left[-\frac{Ai(x_{>}-2E)Ai(-x_{<}-2E)}{Ai(-2E)Ai'(-2E)}\right]\delta'(y)\, dxdy\nonumber \\
\!\!&\!=\!&\!\! \frac{Ai'(-2E)Ai'(-2E)}{Ai(-2E)Ai'(-2E)}=\frac{Ai'(-2E)}{Ai(-2E)},
\label{(4.8)}
\end{eqnarray}
which is clearly divergent. However, since the difference of the two divergent series in the denominator is convergent, (\ref{(4.7)}) can be rewritten as
\be\label{(4.10)}
\beta=\frac{Ai(0)Ai(-2E)}{Ai(0)Ai'(-2E)-Ai'(0)Ai(-2E)},
\ee 
 
\begin{figure}[ht]
\centering
\includegraphics[width=8cm]{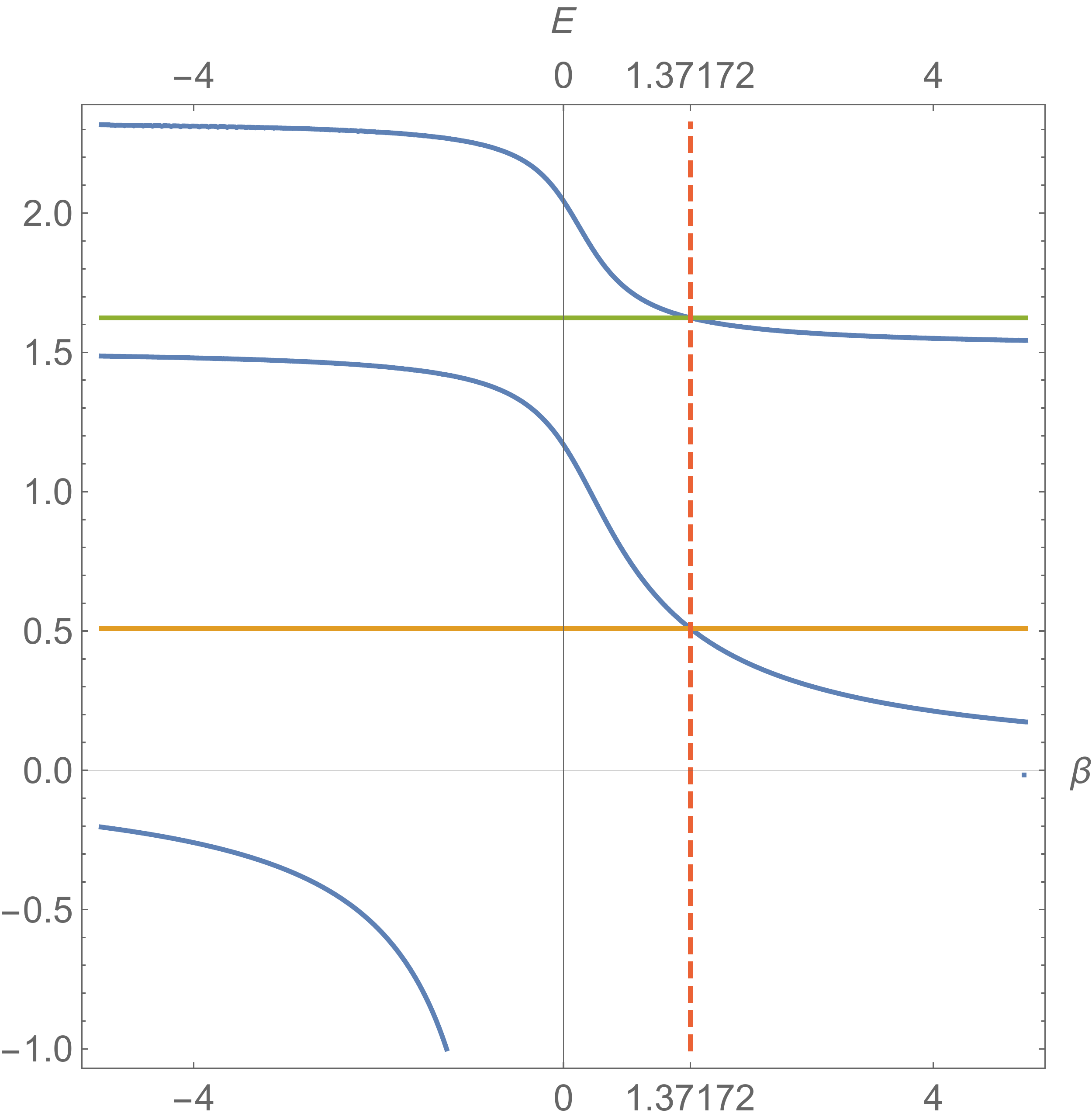}
\caption{\small The level crossings occurring at $\beta_{0}=1.37172$  between the eigenenergy $E_{2}(\beta)$ (middle blue curve) and  $E_{1}(0)$ (orange horizontal line), as well as between the eigenenergy $E_{4}(\beta)$ (upper blue curve) and  $E_{3}(0)$  (green horizontal line).
\label{fig8}}
\end{figure}

Due to the very nature of all the unperturbed odd eigenvalues (see again \cite{LM}, in particular Theorem 3.5 and Corollary 3.6), $Ai'(-2E _{2n-1})=0$. Hence, the special value of the coupling given by
\be\label{(4.11)}
\beta_{0}=-\frac{Ai(0)}{Ai'(0)},
\ee 
is exactly the location of all the level crossings occurring between each even eigenvalue of $H_{\beta}$ (pertaining to an antisymmetric bound state) and the next lower unperturbed odd eigenvalue (pertaining to a symmetric bound state), as shown in Figure~\ref{fig8}. Furthermore, the discrete spectrum of $H_{\beta_{0}}$ consists of doubly degenerate eigenvalues since $E_{2n}(\beta_{0})=E_{2n-1}(0)=\lambda_{2n-1}$ for any $n\geq 1$.

The above findings are clearly in perfect agreement with those of \cite{AFR1} in which the free Hamiltonian was that of the quantum harmonic oscillator. As a matter of fact, whilst in that paper (see also \cite{Albeverio16}) the location of all the level crossings was
\begin{eqnarray}
\beta_{0}=\frac{1}{(H_{0}+1) ^{-1}(0,0)} 
=\frac{1}{\displaystyle\frac{1}{\pi^{1/2}}\sum _{n=0}^{\infty }\displaystyle \frac{H ^{2}_{2n}(0)}{2^{2n}(2n)!(2n+3/2)}}=\frac{\Gamma(1/4)}{2\Gamma(3/4)}\approxeq 1.47934,
\label{(4.12)}
\end{eqnarray}
here we have
\be\label{(4.13)}
\beta_{0}=\frac{1}{(H_{0}) ^{-1}(0,0)}=\frac{2}{\displaystyle\sum _{n=1}^{\infty}\frac{1}{E^{2}_{2n-1}}}=-\frac{Ai(0)}{Ai'(0)}\approxeq 1.37172,
\ee 

The figure also shows that, due to the level crossing phenomenon, the curves representing the perturbed eigenvalues no longer have the unperturbed ones as horizontal asymptotes, differently from the level rearrangement pattern (see \cite{FF,AFR1,AlbeverioFa16}). This fact further confirms the extremely singular nature of the nonlocal $\delta'$ interaction.

In analogy with what was seen in \cite{AFR1,Albeverio16} in the case of the harmonic oscillator perturbed by the $\delta'$-interaction, the symmetry of the ground state wave function changes from symmetric to antisymmetric crossing the point $\beta=\beta_{0}$. Hence, the latter critical value is where a quantum phase transition takes place.  
\medskip

\noindent
\textbf{Remark 4.2} It might be worth pointing out that the level crossings observed in the spectrum of the three-dimensional isotropic harmonic oscillator perturbed by a $\delta$-interaction occur at the point (see  \cite{AFR1,AlbeverioFa16,FI2,AlbeverioFaRi16}) 
\be\label{(4.14)}
\beta_{0}=\frac{\pi\Gamma(1/4)}{\Gamma(3/4)}\approxeq 9.29495 ,
\ee
However, as stressed in the above-mentioned papers, the three-dimensional crossings are of an opposite nature: as $\beta>\beta_{0}$ (strongly attractive perturbation), the energy of each new bound state created by the point perturbation, clearly symmetric as the value of the total angular momentum is even, falls below that of the next lower antisymmetric bound state.

\subsection{Local $\delta'$ interaction: self-adjoint extensions approach}

In order to illustrate the sensitivity of these calculations to the precise interpretation of  highly singular point potentials we shall examine the effect of the potential $-a \delta(x)+b \delta'(x)$  on the conical oscillator when interpreted,
as is frequently done, as the set of boundary conditions \cite{GGN}.

It is noteworthy that, while the nonlocal $\delta'(x)$ interaction is not compatible with a Dirac delta interaction $-\lambda\,\delta(x)$ due to the incompatibility of the matching conditions defining each one, this is not the case for the local $\delta'(x)$ interaction.  In fact, a potential of the form $-a\,\delta(x)+b\,\delta'(x)$ may be defined through matching conditions (1.3). In this case, $b\,\delta'(x)$ has a local character. Then, the total Hamiltonian $H=H_0-a\,\delta(x)+b\,\delta'(x)$ is self-adjoint on a domain of functions showing a discontinuity at the origin, so that the product of each function $\psi(x)$ in this domain are given by, respectively:
\begin{eqnarray}
\delta(x)\psi(x)&=&\frac{\psi(0+)+\psi(0-)}{2}\delta(x)\\
\delta'(x)\psi(x)&=&\frac{\psi(0+)+\psi(0-)}{2}\delta'(x)-\frac{\psi'(0+)+\psi'(0-)}{2}\delta(x).
\end{eqnarray}
A slightly generalisation of these products was given in \cite{ZOLOTARYUK,Semitransparent}.

The previous analysis relative to the characterization of the resolvent of the operator $H_0+\lambda\,\delta'(x)$, for nonlocal $\delta'(x)$ is not straightforwardly applicable here.  Instead, we use a technique that has been discussed in \cite{GGN}, which relies on calculations with the Green functions of $H_0$ and $H$. In fact, if $G_0(x,x',E)$ is the Green function of $H_0$, then bound states $\psi(x)$ with energy $E$ corresponding to the total Hamiltonian $H$ as above have the following form:
\begin{equation}
\psi(x)=\int G_0(x,x',E) [-a\,\delta(x')+b\,\delta'(x')]\,\psi(x')\,dx'\,.
\end{equation}
This expression yields an homogeneous system of for equations with four unknowns. In order to obtain non-trivial solutions, the determinant of the system has to be equal to zero. This determinant can be further simplified so as to obtain the following expression \cite{GGN}:
\begin{equation}\label{deterlarry}
\left|\begin{array}{ccc}
2  &  -1-b  &  0\\[1ex]
-1 &\displaystyle 1+\frac{a}{2}A+\frac{b}{2} & \displaystyle\frac{b}{2} A\\ [1ex]
0 & \displaystyle -\frac{b}{A} &1
\end{array}
\right|=1+a A+b^2=0\,.
\end{equation}
Then,  the energy levels of the total Hamiltonian $H$ are the roots of the above determinant.

\begin{figure}[ht]
\centering
\includegraphics[width=8cm]{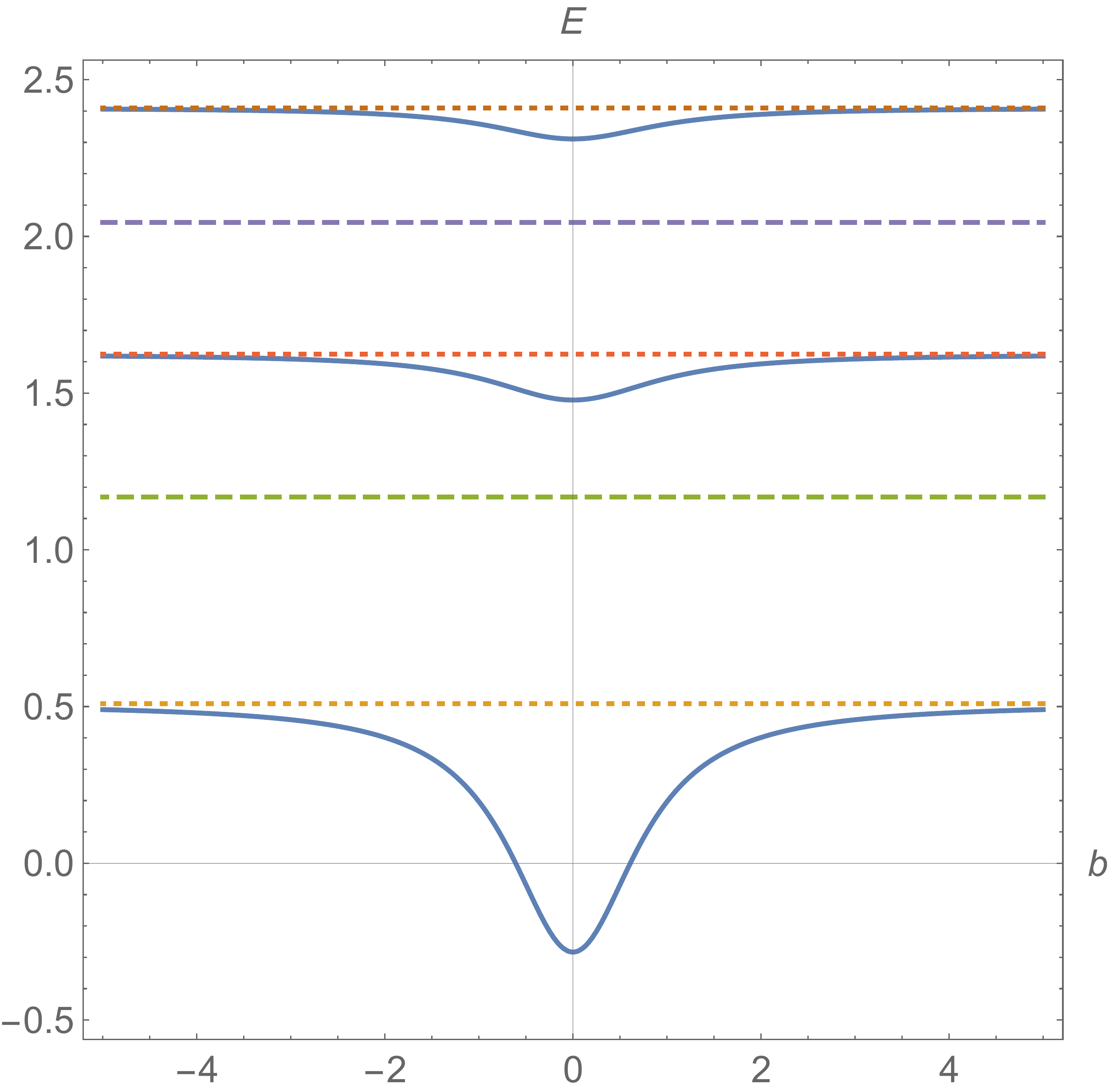}
\caption{\small The first five energy levels of the conic potential plus the perturbation $-a\delta(x)+b\delta'(x)$ for $a=1$ as functions of $b$: in blue the symmetric levels and dashed the antisymmetric ones (they do not depend on $b$). The dotted lines correspond to the symmetric energy levels of the unperturbed conic potential.
\label{fig9}}
\end{figure}

Here, $A$ is  related to the Green function $G_0(x,x';E)$  defined in (\ref{(1.2)})  by
\begin{equation}
A=G_0(0+,0;E)=\frac{Ai(-2E)}{Ai'(-2E)},
\end{equation}
Therefore the vanishing of the determinant (\ref{deterlarry}) reduces to
\begin{equation}\label{zxxr}
\frac{Ai'(-2E)}{Ai(-2E)} =-\frac{a}{b^2+1}.
\end{equation}
Hence, if $a=0$ there are no new states, and if $b=0$ the new states coincide with those of (\ref{(2.9)}) with $a=\lambda$. In Figure~\ref{fig9} we plot the structure of the first five energy levels for this Hamiltonian for a particular value of $a$, as functions of $b$.

\section{Final remarks}

In this note we have shown that the self-adjoint Hamiltonian $H_0:=\frac{1}{2} \left[-\frac{d^{2} }{dx^{2} } + \left|x \right| \right]$, describing a quantum oscillator with conical confinement in place of the standard harmonic one, can be decorated with a point impurity of either the $\delta$-type, or the local $\delta'$-type or even the nonlocal $\delta'$-type to yield an exactly solvable model in all three cases, that is to say the discrete spectra of the perturbed operators can be explicitly obtained by solving transcendental equations involving the Airy function and its derivative.

The study of the spectra can be done either by constructing directly the resolvent operator of the total Hamiltonian or by analysing its Green function. Since the Green function is the integral kernel of the resolvent operator, both methods are equivalent. 

We have obtained some results that are in accordance with those obtained when the unperturbed Hamiltonian $H_0$ is the one-dimensional harmonic oscillator. These are:

\begin{itemize}
\item[(i)] 
When $H_0$ undergoes a perturbation of the type $-\lambda\,\delta(x)$, even energy levels (pertaining to antisymmetric bound states) are unaffected by the perturbation. This is not the case for odd energy levels (pertaining to symmetric bound states) for which the energy decreases as $\lambda$ increases. For odd energy levels excluding the first one, the function of the energy $E$ in terms of $\lambda$ show an asymptotic behaviour towards the next and previous even energy values as $\lambda\to -\infty$ and $\lambda\to\infty$, respectively.  The first energy levels shows the same behaviour for $\lambda\to-\infty$, while the value of the energy goes to $-\infty$ as $\lambda\to\infty$.

\item[(ii)] 
However, when the perturbation is of the type $-\lambda\,\delta(x-x_0)$, even levels are no longer constant with $\lambda$, but instead undergo a change, which is usually small. All levels decrease as $\lambda$ increases and the energy of the first one still goes to $-\infty$ as $\lambda\to\infty$.

\item[(iii)] 
When the perturbation is of the type $\delta'(x)$ the situation changes dramatically. First of all, we consider two types of $\delta'(x)$ interactions: nonlocal and local. We have constructed the resolvent operator for the Hamiltonian $H_0$ decorated with a nonlocal $\delta'$ perturbation centred at the origin. Contrary to the case of the Dirac delta perturbation, this construction requires renormalization. Odd energy levels (pertaining to symmetric bound states) are unchanged with the renormalized form factor $\beta$, but even energy levels (pertaining to antisymmetric bound states) decrease as $\beta\to\infty$. At a particular value of $\beta$, which is the same for all levels $\beta=1.37172$), we observe a noteworthy phenomenon: the existence of level crossings. The energy of each even level becomes lower than the energy of the previous odd level (except the first one) and remains lower as $\beta\to \infty$.

\item[(iv)] 
When the perturbation of type $\delta'(x)$ is local, it is possible to combine it with a Dirac delta perturbation, so that the total Hamiltonian becomes $H=H_0-a\,\delta(x)+b\,\delta'(x)$ in this case. For simplicity, we have investigated the energy levels of $H$ using its Green function. While the energy of even levels remains unaffected, for the odd levels undergo a lowering without level crossings, provided that $a\ne 0$. If $a=0$, the $b\,\delta'(x)$ term does not affect to any of the energy values.
\end{itemize}

We believe the current findings pave the way to the investigation of the conic or the pyramidal oscillator perturbed by a point impurity in two and three dimensions. We also believe that it should be possible to show that the self-adjoint operator $H_{\beta}$ defined in Section 5.1 by renormalizing the coupling constant can also be obtained by means of the so-called Cheon-Shigehara approximation involving a suitable triple of Dirac distributions, given that such a result holds when the free Hamiltonian is the one of the harmonic oscillator (see \cite{Albeverio16} and the references therein).

\section*{Acknowledgements}

Partial financial support is acknowledged to the Spanish Junta de Castilla y Le\'on (VA057U16) and
MINECO (Project MTM2014-57129-C2-1-P). S Fassari gratefully acknowledges financial support from the 
``Grants for Visiting Researchers at the Campus of International Excellence Triangular-E3", as part of the ``Attraction of Excellent Researchers and Stays for Visiting Researchers Program", carried out under the subvention of the Ministry of Education, Culture and Sports to the Campus of International Excellence Triangular-E3.  S Fassari and ML Glasser also wish to thank the entire staff at Departamento de F\'{\i}sica Te\'{o}rica, At\'{o}mica y \'{O}ptica, Universidad de Valladolid, for their warm hospitality throughout their stay.

\end{document}